\documentclass[aps,amsmath,amssymb,prl,twocolumn,showpacs,superscriptaddress]{revtex4-1}
\usepackage{bm}
\usepackage{epsfig}
\usepackage[usenames]{color}
\usepackage{graphicx}
\usepackage{amsfonts}

\newcommand{\olcite}[1]{\cite{#1}} 

\begin{document}

\title{Quantum Fluctuations and Dynamic Clustering of Fluctuating Cooper Pairs}
\author{Andreas Glatz}
\affiliation{Materials Science Division, Argonne National
Laboratory, 9700 S.Cass Avenue, Argonne Il 60439}
\author{A.~A.~Varlamov}
\affiliation{Materials Science Division, Argonne National
Laboratory, 9700 S.Cass Avenue, Argonne Il 60439}
\affiliation{CNR-SPIN,
Viale del Politecnico 1, I-00133 Rome, Italy}
\author{V.~M.~Vinokur}
\affiliation{Materials Science Division, Argonne National
Laboratory, 9700 S.Cass Avenue, Argonne Il 60439}

\date{\today }

\begin{abstract}
We derive exact expressions for the fluctuation conductivity in two dimensional superconductors
as a function of temperature and magnetic field in the whole fluctuation region above the
upper critical field $H_{\mathrm{c2}}\left( T\right)$.  Focusing on the vicinity of the quantum phase
transition near zero temperature,  we propose that as the
magnetic field approaches the line near $H_{\mathrm{c2}}\left(
0\right)$ from above, a peculiar dynamic state consisting of clusters of coherently
rotating fluctuation Cooper-pairs forms and estimate the characteristic
size and lifetime of such clusters.  We find the zero temperature magnetic field dependence of the the transverse
magnetoconductivity above $H_{\mathrm{c2}}\left( 0\right) $ in layered superconductors.
\end{abstract}

\maketitle

The understanding of the mechanisms of superconducting fluctuations, achieved
during the past decades~\cite{LV09} gave a unique tool providing the information about the
microscopic parameters of superconductors.  The fluctuations are customary described in terms of
the so-called quantum corrections to conductivity, i.e. Aslamazov-Larkin (AL)~\cite{AL68} and
Maki-Thompson (MT)~\cite{M68,T70} corrections, and/or the fluctuation corrections to 
the density of states (DOS) of the normal excitations~\cite{ILVY93,BDKLV93}.  The classical results
obtained first in the vicinity of the superconducting critical temperature $T_c$ were
generalized to the far from $T_{\mathrm c0}$~\cite{AV80,L80,ARV83} and relatively high 
fields~\cite{Ab85} regions.  More recently, quantum fluctuations (QF) entered the
focus. Effects of QFs on magnetoconductivity and magnetization of 2D SCs were
studied at low temperatures and fields close to $H_{\mathrm{c2}}\left(
0\right) $ in Ref.~\olcite{GL01}. It was found~\cite{BEL00} that in
granular SCs at very low temperatures and close to $H_{\mathrm{c2}}\left(
0\right) $, the  AL contribution is $\propto T^{2}$, and the magnetoresistance
grows due to the suppression of the DOS. 
The study of different fluctuation contributions to the
Nernst-Ettingshausen effect in 2D SCs, in a wide region above the transition
line $H_{\mathrm{c2}}\left( T\right) $, demonstrated the importance of renormalization of the
diffusion coefficient (DCR) due to QF~\cite{SSVG09}.
Yet, the region near $T=0$
and magnetic fields near $H_{\mathrm{c2}}\left( 0\right)$ remains understudied and what more,
the universal picture combining QF at high magnetic fields and conventional finite temperature
quantum corrections is still lacking.

In this paper we offer a general unifying description of fluctuation-induced
conductivity of a disordered 2D SC in a perpendicular magnetic field that
holds everywhere above the transition line $H_{\mathrm{c2}}\left( T\right)$.
Considering the vicinity of $T=0$, we find
that on the approach to $H_{\mathrm{c2}%
}\left( 0\right) $ from the above, a peculiar dynamic state forms comprised by the
clusters of coherently rotating fluctuation Cooper-pairs and estimate
the characteristic size $\xi _{\mathrm{QF}}\left( H\right) $ and lifetime $%
\tau _{\mathrm{QF}}\left( H\right) $ of such clusters. 
Using the derived values of $\xi _{\mathrm{QF}}$ and $\ \tau _{\mathrm{QF}}$
we cross-check our conclusions by reproducing the  results of~\olcite{BEL00,GL01,SSVG09}.
%
\begin{figure}[t]
\includegraphics[width=.95\columnwidth]{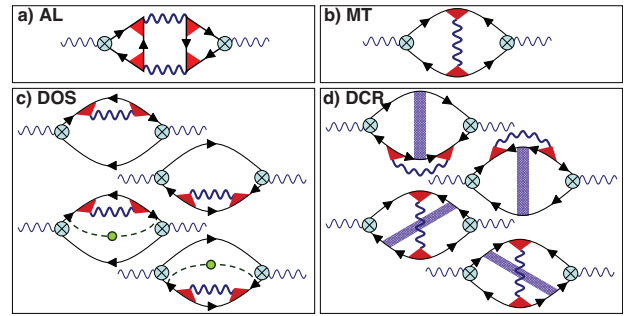}
\caption{(Color online) Feynman diagrams for the leading-order contributions
to the electromagnetic response operator. Wavy lines stand for
fluctuation propagators, solid lines with arrows are impurity-averaged
normal state Green's functions, crossed circles are electric field vertices,
dashed lines with a circle represent additional impurity renormalizations, and
triangles and dotted rectangles are impurity ladders accounting for the electron
scattering on impurities (Cooperons). }
\label{conddia}
\end{figure}
%

We start with the derivation of the universal expressions for fluctuation contribution to
conductivity and DOS.  The electromagnetic response operator in the
leading order is defined by ten diagrams~\cite{LV09} shown in Fig. \ref{conddia}.
The corresponding contributions to conductivity at arbitrary magnetic fields
and temperatures $T_{\mathrm{c}}\left( H\right) <T\ll \tau ^{-1}$ ($\tau $
is the electron elastic scattering time) are:

\begin{widetext}%
\begin{eqnarray}
\delta\sigma_{xx}^{\mathrm{AL}}  & =&\frac{e^{2}}{\pi}\sum_{m=0}^{\infty}%
(m+1)\int_{-\infty}^{\infty}\frac{dx}{\sinh^{2}\pi x}\left\{  \frac
{\operatorname{Im}^{2}\mathcal{E}_{m}}{\left\vert \mathcal{E}_{m}\right\vert
^{2}}+\frac{\operatorname{Im}^{2}\mathcal{E}_{m+1}}{\left\vert \mathcal{E}%
_{m+1}\right\vert ^{2}}+\frac{\operatorname{Im}^{2}\mathcal{E}_{m+1}%
-\operatorname{Im}^{2}\mathcal{E}_{m}}{\left\vert \mathcal{E}_{m}\right\vert
^{2}\left\vert \mathcal{E}_{m+1}\right\vert ^{2}}\operatorname{Re}\left[
\mathcal{E}_{m}\mathcal{E}_{m+1}\right]  \right\}  ,\label{1}\\
\delta\sigma_{xx}^{\mathrm{MT}}  & =&\frac{e^{2}}{\pi}\left(  \frac{h}{t}\right)
\sum_{m=0}^{M}\underbrace{\frac{{1}}{\gamma_{\phi}+\frac{2h}{t}\left(
m+1/2\right)  }\int_{-\infty}^{\infty}\frac{dx}{\sinh^{2}\pi x}\frac
{\operatorname{Im}^{2}\mathcal{E}_{m}}{\left\vert \mathcal{E}_{m}\right\vert
^{2}}}_{\delta\sigma_{xx}^{\mathrm{MT(an)}}+\delta\sigma_{xx}^{\mathrm{MT(reg2)}}%
}+\underbrace{\frac{e^{2}}{\pi^{4}}\left(  \frac{h}{t}\right)  \sum_{m=0}%
^{M}\sum_{k=-\infty}^{\infty}\frac{4\mathcal{E}_{m}^{\prime\prime}\left(
t,h,|k|\right)  }{\mathcal{E}_{m}\left(  t,h,|k|\right)  }}_{\delta\sigma
_{xx}^{\mathrm{MT(reg1)}}},\;\label{2}\\
\delta\sigma_{xx}^{\mathrm{DOS}}  & =&\frac{2e^{2}}{\pi^{3}}\left(  \frac{h}%
{t}\right)  \sum_{m=0}^{M}\int_{-\infty}^{\infty}\frac{dx}{\sinh^{2}\pi
x}\frac{\operatorname{Im}\mathcal{E}_{m}\operatorname{Im}\mathcal{E}%
_{m}^{\prime}}{\left\vert \mathcal{E}_{m}\right\vert ^{2}},\text{ and}%
\quad\delta\sigma_{xx}^{\mathrm{DCR}}=\frac{4e^{2}}{3\pi^{6}}\left(  \frac{h}%
{t}\right)  ^{2}\sum_{m=0}^{M}(m+\frac{1}{2})\sum_{k=-\infty}^{\infty}%
\frac{8\mathcal{E}_{m}^{\prime\prime\prime}\left(  t,h,|k|\right)
}{\mathcal{E}_{m}\left(  t,h,|k|\right)  }.\label{3}%
\end{eqnarray}
\end{widetext}

\begin{figure}[b]
\includegraphics[width=.98\columnwidth]{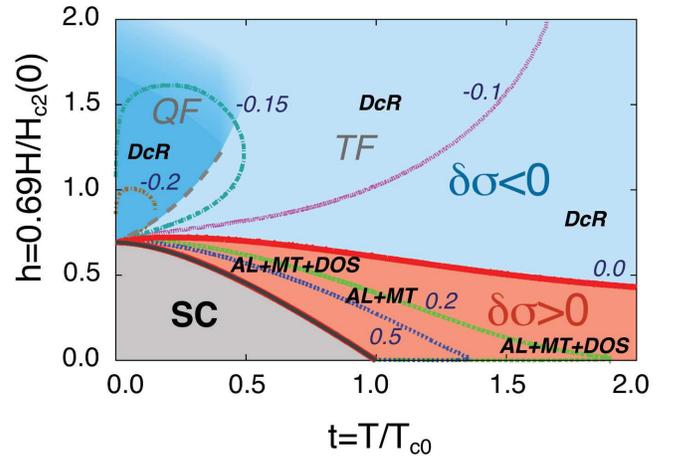}
\caption{(Color online) Contours of constant fluctuation conductivity [$%
\protect\delta\sigma=\delta\sigma _{xx}^{\mathrm{(tot)}}\left( t,h\right) $ shown in units of $e^2$]. The dominant FC contributions are
indicated in bold-italic labels. The dashed line separates the domain of quantum
fluctuations (QF) [dark area of $\protect\delta\sigma>0$] and thermal fluctuations
(TF). The contour lines are obtained from Eqs. (\protect\ref{1})-(\protect
\ref{3}) with $T_{c0}\protect\tau=0.01$ and $T_{c0}\protect\tau_{\protect\phi%
}=10$. }
\label{contours}
\end{figure}

Here $t=T/T_{\mathrm{c0}},$ $h=\frac{\pi ^{2}}{8\gamma _{E}}%
\frac{H}{H_{\mathrm{c2}}\left( 0\right) }=0.69\frac{H}{H_{\mathrm{c2}}\left(
0\right) }$, $\gamma _{E}=e^{\gamma _{e}}$ ($\gamma _{e}$ is the Euler
constant), $M=\left( tT_{\mathrm{c0}}\tau \right) ^{-1},\gamma _{\phi }=\pi
/\left( 8T_{\mathrm{c0}}\tau _{\phi }\right) $, $\tau _{\phi }$ is the
phase-breaking time, $\mathcal{E}_{m}\equiv \mathcal{E}_{m}\left(
t,h,ix\right) $ with $\mathcal{E}_{m}\left( t,h,z\right) =\ln t+\psi \left[
\frac{1+z}{2}+\frac{2h}{t}\frac{\left( 2m+1\right) }{\pi ^{2}}\right] -\psi
\left( \frac{1}{2}\right) $ and its derivatives $\mathcal{E}_{m}^{(p)}\left(
t,h,z\right) \equiv \partial _{z}^{p}\mathcal{E}_{m}\left( t,h,z\right) $.
The sum of Eqs.\thinspace (\ref{1})-(\ref{3}) present the general
expression for the total fluctuation correction to conductivity $\delta\sigma _{xx}^{\mathrm{(tot)}}\left( T,H\right) $ that holds in the complete 
$T$-$H$ phase diagram above the line $H_{c2}(T)$.
We analyzed these expressions
both analytically (by finding the asymptotic expressions for $\delta\sigma _{xx}^{%
\mathrm{(tot)}}$ in nine qualitatively different domains) and numerically
(by developing a program which calculates the complete surface $\delta\sigma
_{xx}^{\mathrm{(tot)}}\left( T,H\right) $ for given parameters $\tau $ and $%
\tau _{\phi }$) \olcite{GVVB10}.  We show that the singular growth of the \textit{conductivity} 
near $T_{\mathrm{c0}}$ transforms into a relatively weak
(logarithmic) growth of fluctuation correction to \textit{resistivity} as
one moves along the $H_{\mathrm{c2}}(T)$ line towards the low-temperatures
region (i.e. close to $H_{\mathrm{c2}}(0)$). The total fluctuation
correction to the conductivity $\delta\sigma ^{\mathrm{(tot)}}$ remains negative
in a wide domain relatively far from the $H_{\mathrm{c2}}\left(
T\right) $ line (see Fig.~\ref{contours} where the regions with the dominating
fluctuation contributions to magneto-conductivity are shown).

The analysis of fluctuation corrections enables us to develop a
qualitative picture of the quantum phase transition in the
vicinity of $H_{\mathrm{c2}}\left( 0\right) $. We start by refreshing
the well-established qualitative description of the
vicinity of $T_{\mathrm{c0}}$.  Here the lifetime of fluctuation-induced Cooper 
pairs (FCP), $\tau _{%
\mathrm{GL}}$, is obtained using the uncertainty principle: $\tau _{\mathrm{%
GL}}\sim \hbar /\Delta E$, where $\Delta E$ is the difference $k_{B}(T-T_{\mathrm{c0}%
})$ ensuring that $\tau _{\mathrm{%
GL}}$ should become infinite below $T_{\mathrm{c0}}$.
This yields the standard Ginzburg-Landau time $\tau _{%
\mathrm{GL}}\sim \hbar /k_{B}(T-T_{\mathrm{c0}})\sim \hbar /\left( k_{B}T_{%
\mathrm{c0}}\epsilon \right)$, where $\epsilon =\left( T-T_{\mathrm{c0}%
}\right) /T_{\mathrm{c0}}\ll 1$ is the reduced temperature. The 
coherence length $\xi _{\mathrm{GL}}\left( T\right) $ is estimated
as the distance which two electrons move apart during the GL time: $\xi _{%
\mathrm{GL}}\left( \epsilon \right) =\left( \mathcal{D}\tau _{\mathrm{GL}%
}\right) ^{1/2}\sim \xi _{\mathrm{BCS}}/\sqrt{\epsilon }$ ($\xi _{\mathrm{BCS%
}}$ is the BCS coherence length, $\mathcal{D}$\ is the diffusion coefficient).
The order parameter $\Delta ^{\mathrm{(fl)}}\left( \mathbf{r},t\right) $
varies on the larger scale $\xi _{\mathrm{GL}}\left(
T\right) \gg \xi _{\mathrm{BCS}}$.  The ratio of the FCP concentration to the
corresponding effective mass $n_{\mathrm{c.p.}}/m_{\mathrm{c.p.}}\sim \xi _{%
\mathrm{GL}}^{2-D}\left( \epsilon \right) $ (with the logarithmic accuracy) 
in 2D is constant~\cite{LV09}.

The principal fluctuation contributions to conductivity are positive and originate from
 direct FCP charge transfer (AL contribution) $\delta\sigma _{xx}^{\mathrm{AL}%
}\sim \left( n_{\mathrm{c.p.}}/m_{\mathrm{c.p.}}\right) e^{2}\tau _{\mathrm{%
GL}}\sim e^{2}/\epsilon $ and the specific quantum process of the
one-electron charge transfer related to the coherent scattering of electrons on elastic impurities
forming FCPs  (anomalous MT contribution) $\delta\sigma _{xx}^{%
\mathrm{MT(an)}}\sim \frac{e^{2}}{\epsilon }\ln \left( \epsilon /\gamma
_{\phi }\right) $ \cite{LV09} [Fig.~\ref{conddia}a,b)]. However, these two
contributions do not capture the complete effect of fluctuations on
conductivity. The involvement of quasi-particles in the fluctuation pairing
results in the opening of the pseudo-gap in the one-electron spectrum. The
lack of corresponding electrons at the Fermi level leads to a decrease of
the one-electron Drude-like conductivity. Such an indirect fluctuation
effect formally is described by the third group of four diagrams in Fig. \ref{conddia}c),
which are usually called DOS diagrams. The corresponding
contribution has the opposite sign with respect to the AL and MT
contributions, but close to $T_{\mathrm{c0}}$ turns to be less singular as a
function of temperature\cite{LV09}:
\begin{equation}
\delta\sigma _{xx}^{\mathrm{DOS}}\sim -\frac{2n_{\mathrm{c.p.}}e^{2}\tau }{m_{%
\mathrm{e}}}\sim -e^{2}\int \frac{\xi _{\mathrm{BCS}}^{2}d^{2}\mathbf{q}}{%
\epsilon +\xi _{\mathrm{BCS}}^{2}q^{2}}\sim -e^{2}\ln \frac{1}{\epsilon }.
\label{DOS}
\end{equation}%
The DOS contribution formally appears due to the summation of \ $%
\left\langle \left\vert \Delta ^{\mathrm{\left( fl\right) }}\left( \mathbf{q}%
,\omega \right) \right\vert ^{2}\right\rangle $ over all long-wave-length
fluctuation modes ($q\lesssim \xi _{\mathrm{BCS}}^{-1}\sqrt{\epsilon }$) in
the static approximation ($\omega \rightarrow 0$). Relatively far from $T_{%
\mathrm{c0}}$, $\delta\sigma _{xx}^{\mathrm{DOS}}$ can change the sign of $%
\delta\sigma _{xx}^{\mathrm{(tot)}}$, but this happens only in the case of strong
pair-breaking, when the phase-sensitive anomalous MT contribution is
suppressed. Finally, the last four diagrams in Fig. \ref{conddia}d) together
with the regular part of the MT diagram describe the renormalization of the
diffusion coefficient (DCR) in the presence of fluctuation pairing. Close to
$T_{\mathrm{c0}}$ their contribution is not singular, but becomes of
primary importance at very low temperatures.

At zero temperature and the field above $H_{\mathrm{c2}}\left( 0\right) $, the
systematics of the fluctuation contributions to the conductivity
considerably changes with respect to that close to $T_{\mathrm{c0}}$. The
collisionless rotation of FCPs (they do not \textquotedblleft feel" the
presence of elastic impurities) results in the lack of their direct
contribution to the longitudinal (along the applied electric field) electric
transport (analogously to the suppression of one-electron conductivity $\delta\sigma
_{xx}^{\left( \mathrm{e}\right) }$ in strong magnetic fields $\omega
_{c}\tau \gg 1$, see Ref.~\olcite{A88})  and the AL contribution to $\delta\sigma
_{xx}^{\left( \mathrm{tot}\right) }$ becomes zero. The anomalous MT and DOS
contributions turn zero as well but due to different reasons: Namely, the former
vanishes since magnetic fields as large as $H_{\mathrm{c2}}\left( 0\right) $
completely destroy the phase coherence, whereas the latter disappears since magnetic
field suppresses the fluctuation gap in the one-electron spectrum. Therefore
the effect of fluctuations on the conductivity at zero temperature is
reduced to the renormalization of the one-electron diffusion coefficient.
FCPs here occupy the lowest Landau level, but all the dynamic fluctuations
in the interval of frequencies from $0$ to $\Delta _{\mathrm{BCS}}$ should
be taken into account. The corresponding fluctuation propagator at zero
temperature close to $H_{\mathrm{c2}}\left( 0\right) $ has the form $%
L_{0}\left( \omega \right) =N_{0}^{-1}\left( \widetilde{h}+\omega /\Delta _{%
\mathrm{BCS}}\right) ^{-1}$ and
\begin{equation}
\delta\sigma _{xx}^{\mathrm{DCR}}+\delta\sigma _{xx}^{\mathrm{MT(reg)}}\sim -\frac{e^{2}%
}{\Delta _{\mathrm{BCS}}}\int_{0}^{\Delta _{\mathrm{BCS}}}\frac{d\omega }{%
\widetilde{h}+\frac{\omega }{\Delta _{\mathrm{BCS}}}}\sim -e^{2}\ln \frac{1}{%
\widetilde{h}},  \label{DCR}
\end{equation}%
where $\widetilde{h}=\left[ H-H_{\mathrm{c2}}\left( 0\right) \right] /H_{%
\mathrm{c2}}\left( 0\right) $ is the parameter playing the role of the reduced
temperature $\epsilon$ in the case of the classical transition;  $\Delta _{
\mathrm{BCS}}$ is the BCS value of the gap at zero temperature in zero
field, and $N_{0}$ is the one-electron density of states.

The microscopic theory provides (unlike the GL approach) a correct
description of the short wavelength and high frequency dynamic fluctuations
for the complete magnetic field and temperature range. For instance, in the region
of QFs, but for non-zero temperatures $t\ll \widetilde{h}$ (see Fig. \ref{contours})
the AL and the anomalous MT contributions are equal to each other
and grow as the square of temperature; moreover, one of them is exactly
cancelled by the appearing negative contribution of the four DOS-like
diagrams:
\begin{equation}
\delta\sigma _{xx}^{\mathrm{AL}}=\delta\sigma _{xx}^{\mathrm{MT(an)}}=-\delta\sigma _{xx}^{%
\mathrm{DOS}}=\frac{4e^{2}\gamma _{E}^{2}t^{2}}{3\pi ^{2}\widetilde{h}^{2}}.
\label{ALlow}
\end{equation}

While Eq. (\ref{DOS}) defines the characteristic wavelength $\xi _{%
\mathrm{GL}}\left( T\right) $ of the fluctuation modes close to $T_{\mathrm{%
c0}}$,  Eq. (\ref{DCR}) defines the
characteristic coherence time $\tau _{\mathrm{QF}}\left( \widetilde{h}%
\right) $ of QFs near $H_{\mathrm{c2}}\left( 0\right)$ (where $t\ll
\widetilde{h}$ ). The integral there is determined by its lower cut off $%
\omega _{\mathrm{QF}}\sim \Delta _{\mathrm{BCS}}\widetilde{h}$, and the
corresponding time scale is $\tau _{\mathrm{QF}}\sim \hbar \left( \Delta _{%
\mathrm{BCS}}\widetilde{h}\right) ^{-1}$. One sees that the functional form of $\tau _{%
\mathrm{QF}}$  is analogous to that of $\tau _{\mathrm{GL}}$, this 
can be obtained also from the uncertainty principle. The 
energy, characterizing the proximity to $H_{\mathrm{c2}}\left( 0\right) $%
, is $\Delta E=\hbar \omega _{c}\left(
H\right) -\hbar \omega _{c}\left( H_{\mathrm{c2}}\left( 0\right) \right)
\sim \Delta _{\mathrm{BCS}}\widetilde{h}$ in this case.
However, the spatial scale of QFs close to $H_{\mathrm{c2}}\left( 0\right) $ 
cannot be found from the propagator describing
QFs, since the latter in
the Landau representation does not contain space variables.
Nevertheless, the spatial coherence scale can be estimated from
the value of $\tau _{\mathrm{QF}}$ analogously to consideration near
$T_{\mathrm{c0}}$.  Namely, two electrons with the coherent phase
starting from the same point get separated by 
the distance $\xi _{\mathrm{QF}}\sim \left( \emph{D}\tau _{\mathrm{QF}%
}\right) ^{1/2}\sim \xi _{\mathrm{BCS}}/\sqrt{\widetilde{h}}$ after the time $\tau _{\mathrm{QF}}$.
During this time they participate in multiple fluctuating Cooper pairings
maintaining their coherence.

To clarify the physical meaning of $\tau _{\mathrm{QF}}$ and $\xi _{\mathrm{%
QF}}$, note that near the quantum phase transition at zero temperature, where
$H\rightarrow H_{\mathrm{c2}}\left( 0\right) $, the fluctuations of the
order parameter $\Delta ^{\mathrm{(fl)}}\left( \mathbf{r},t\right) $ become
highly inhomogeneous, contrary to the situation near $T_{\mathrm{c0}}$.
Indeed, below $H_{\mathrm{c2}}\left( 0\right) $, the spatial distribution of
the order parameter at finite magnetic field reflects the existence of
Abrikosov vortices with average spacing (close to $H_{\mathrm{c2}}\left(
0\right) $ but in the region where the notion of vortices is still adequate)
equal to $\xi _{\mathrm{BCS}}/\sqrt{H_{\mathrm{c2}}\left( 0\right) /H}$.
Therefore, one expects that close to and above $H_{\mathrm{c2}}\left(
0\right) $ the fluctuation order parameter also varies over the scale of\ $%
\xi _{\mathrm{BCS}}$. At fields higher but not too far from $H_{\mathrm{c2}%
}\left( 0\right) $ one can expect that these \textquotedblleft vortex-like"
spatial inhomogeneities are preserved over the time scale $\tau _{\mathrm{QF}%
}$, whereas $\Delta ^{\mathrm{(fl)}}\left( \mathbf{r},t\right) $ varies over
the spatial scale $\xi _{\mathrm{BCS}}$. The coherence length $\xi _{\mathrm{%
QF}}$ is thus a characteristic size of a cluster of
coherently rotating FCPs, and $\tau _{\mathrm{QF}}$ estimates the
lifetime of such flickering clusters. 

In terms of these introduced QF characteristics $\tau _{\mathrm{QF}}$ and $\xi _{%
\mathrm{QF}}$ one can understand the meaning of already found microscopic
QF contributions and derive others which are required. For example, the
physical meaning of Eq. (\ref{ALlow}) can be understood as follows: one
could estimate the FCP conductivity by mere replacing $\tau _{\mathrm{GL}%
}\rightarrow \tau _{\mathrm{QF}}$ in the classical AL formula, which would
give $\delta\widetilde{\sigma }^{AL}\sim e^{2}\tau _{\mathrm{QF}}.$ Nevertheless,
as we already noticed, the FCPs at zero temperature cannot move in the
electric field but only rotate. As temperature deviates from zero, the FCP
can change their state due to the interaction with the thermal bath, i.e.
their hopping to an adjacent rotation trajectory along the applied electric
field becomes possible. This means that they can participate in
longitudinal charge transfer. This process can be mapped onto the
paraconductivity of a granular superconductors~\cite{LVV08} at temperatures
above $T_{\mathrm{c0}}$, where the FCP tunnelling between grains occurs in
two steps: first one electron jumps,  then the second follows. The
probability of each hopping event is proportional to the inter-grain
tunneling rate $\Gamma .$ To conserve the superconductive coherence between
both events, the latter should occur during the FCP lifetime $\tau _{\mathrm{%
GL}}.$\ The probability of FCP tunnelling between two
grains is determined as the conditional probability of two one-electron
hopping events and is proportional to $W_{\Gamma }=\Gamma ^{2}$\ $\tau _{%
\mathrm{GL}}.$ \ Coming back to the situation of FCPs above $H_{\mathrm{c2}%
}\left( 0\right) $, one can identify the tunnelling rate with temperature $T$
while $\tau _{\mathrm{GL}}$ corresponds to $\tau _{\mathrm{QF}}.$ Therefore,
in order to get a final expression, $\delta\widetilde{\sigma }^{\mathrm{AL}}$ should
be multiplied by the probability factor $W_{\mathrm{QF}}=t^{2}\tau _{\mathrm{%
QF}}$ of the FCP hopping to the neighboring trajectory: $\delta\sigma _{xx}^{%
\mathrm{AL}}\sim \delta\widetilde{\sigma }^{AL}W_{\mathrm{QF}}\sim e^{2}t^{2}/%
\widetilde{h}^{2},$ which corresponds to the low temperature asymptotic Eq. (\ref{ALlow}).

\begin{figure}[t]
\includegraphics[width=.95\columnwidth]{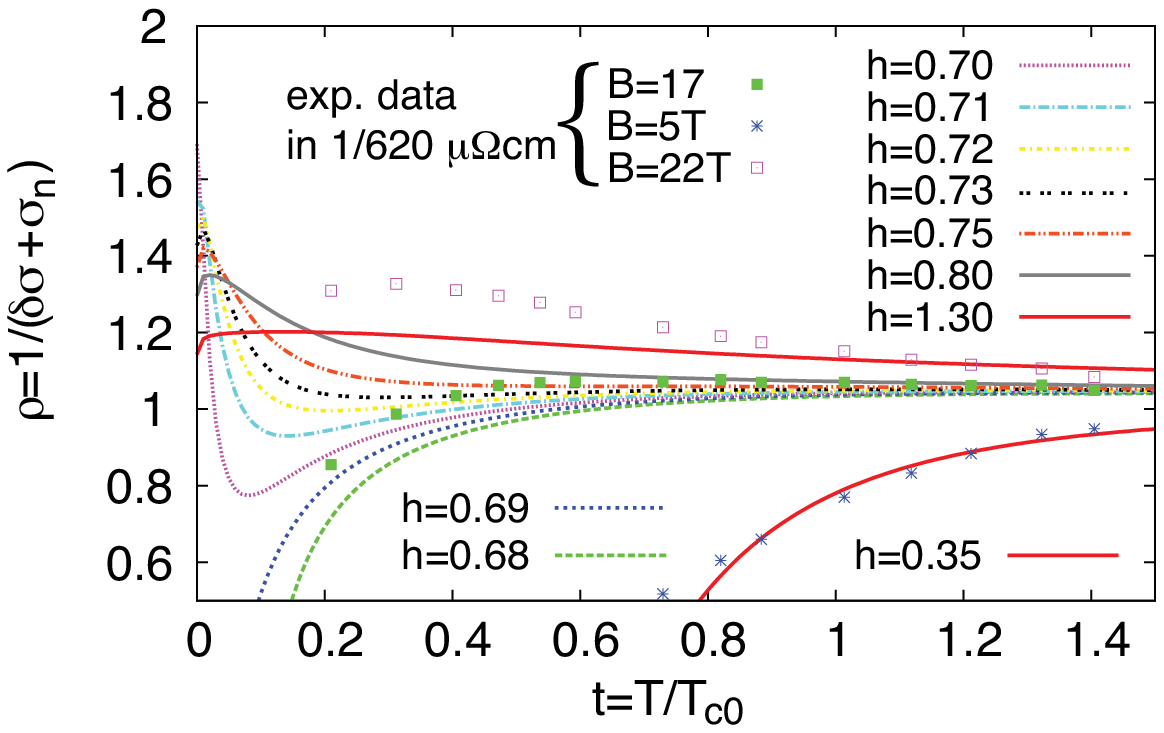}
\caption{(Color online) Temperature dependence of the FC at different fields close to $%
H_{c2}(0)$ and comparison to experimental data for thin films of La$_{2-x}$Sr%
$_{x}$CuO$_{4}$ with $T_{c0}\approx 19$K and $B_{c2}(0)\approx 15$T.~%
\protect\footnote{%
Data is courtesy of B. Leridon, see also Ref.~\olcite{Bridgitte}.}
Note, that for the theoretical curves a fixed $T_{c0}\protect\tau _{\protect%
\phi }=10$ is used which does not necessarily agree with the experimental
value. Nevertheless, the overall behavior can be captured by this rough
comparison. All curves are numerically calculated with $T_{c0}\protect\tau%
=0.01$. }
\label{sigmahconst}
\end{figure}

The total fluctuation contribution to conductivity $\delta\sigma _{xx}^{\left(
\mathrm{tot}\right) }$ at $t\ll \widetilde{h}$  turns out to be negative
and diverges logarithmically when the magnetic field approaches $%
H_{c2}\left( 0\right) :$%
\begin{equation}
\delta\sigma _{xx}^{\mathrm{tot}}=-\frac{2e^{2}}{3\pi ^{2}}\ln \frac{1}{\widetilde{%
h}}-\frac{6\gamma _{E}e^{2}}{\pi ^{2}}\frac{t}{\widetilde{h}}+O\left[ \left(
\frac{t}{\widetilde{h}}\right) ^{2}\right] .  \label{tot}
\end{equation}%
The numerical analysis of Eqs.~(\ref{1})-(\ref{3}) shows that the asymptotic 
Eq.~(\ref{tot}) is valid only within the narrow field range $\widetilde{h}%
\lesssim 10^{-6}$. The nontrivial fact following from Eqs.~(\ref{1})-(\ref{3}%
) is that an increase of  temperature at a fixed value of the magnetic
field in this domain results in a non-monotonic behavior of the fluctuation
part of the conductivity~\cite{GL01}  (see Fig.\thinspace \ref{sigmahconst}),
observed in experiments~\cite{Bridgitte,Batur}.

Now we estimate the contributions of QFs to different
characteristics of the SC in the vicinity of $H_{\mathrm{c2}}\left( 0\right) $.
Applying the Langevin formula $ \chi =e^{2}\left( n/mc\right) \left\langle
R^{2}\right\rangle $ to a coherent cluster and taking $n_{\mathrm{c.p}.}/m_{%
\mathrm{c.p}.}\sim 1$, $R^{2}\sim \xi _{\mathrm{QF}}^{2}\left( \widetilde{h}%
\right) $ one finds the fluctuation magnetic susceptibility $\chi ^{\mathrm{AL}}\sim
\xi _{\mathrm{BCS}}^{2}/c\widetilde{h}$ in an agreement with 
Ref.~\olcite{GL01}. One further reproduces the contribution of QF to the
Nernst coefficient~\cite{SSVG09}.  Using $\nu \sim \left[
\sigma /\left( ne^{2}c\right) \right] d\mu /dT$ and identifying the chemical
potential of FCP, $\mu _{\mathrm{FCP}}$, as $\hbar \omega _{\mathrm{c}%
}\left( H_{\mathrm{c2}}\left( 0\right) \right) -\hbar \omega _{\mathrm{c}%
}\left( H\right) $ (as in Ref.~\olcite{SSVG09}, close to $T_{\mathrm{c0%
}}$, $\mu _{\mathrm{FCP}}=T_{\mathrm{c0}}-T$), one finds that $d\mu _{%
\mathrm{FCP}}/dT\sim dH_{\mathrm{c2}}\left( T\right) /dT\sim -T/\Delta _{%
\mathrm{BCS}}$ and $\nu ^{\mathrm{AL}}\sim \xi _{\mathrm{BCS}}^{2}t/%
\widetilde{h}$.
One can also predict the non-monotonic dependence of the transverse
fluctuation conductivity $\delta\sigma _{zz}^{\left( \mathrm{tot}\right) }\left(
\widetilde{h},0\right) $ above $H_{\mathrm{c2}}\left( 0\right) $ for layered
SCs in a field perpendicular to the layers. The motion of FCPs along the
z-axis has hopping character, and applying the above mapping procedure one
finds: $\delta\sigma _{zz}^{\mathrm{AL}}\left( \widetilde{h},0\right) \sim
\delta\widetilde{\sigma }^{\mathrm{AL}}\Gamma ^{2}$ $\tau _{\mathrm{QF}%
}=e^{2}\Gamma ^{2}/\widetilde{h}^{2}$. The same holds for the quasi-particle
contributions (accounting for the anisotropy factor~\cite{BDKLV93}), and $%
\delta\sigma _{zz}^{\mathrm{DOS}}\sim -\left( \Gamma /E_{\mathrm{F}}\right)
e^{2}\ln \left( 1/\widetilde{h}\right) $. As a result $\delta\sigma _{zz}^{\mathrm{%
tot}}\sim e^{2}\Gamma ^{2}/\widetilde{h}^{2}-\left( \Gamma /E_{\mathrm{F}%
}\right) e^{2}\ln \left( 1/\widetilde{h}\right) $ explaining the strong
growth of the transversal magneto-resistance in organic superconductors at
the edge of the transition at very low temperatures~\cite{MK10}.

Finally, our results can be used for the analysis of novel studies of superconductors
in ultra-high magnetic fields~\cite{Bridgitte} (see also Fig.~\ref{sigmahconst}), for the separation of quantum
corrections in the vicinity of the superconductor--insulator transition \cite{Batur}, and the precise definition of the critical temperature and extraction of the
temperature dependence of $\tau _{\phi }\left( T,H\right)$.

We thank Yu. Galperin, M.\thinspace\ Kartsovnik, A.\thinspace Koshelev, M.
Norman, and T. \thinspace Baturina for useful discussions. The work was
supported by the U.S. Department of Energy Office of Science under the
Contract No. DE-AC02-06CH11357. A.A.V. acknowledges support of the MIUR
under the project PRIN 2008.

\end{document}